\documentclass[reprint,twocolumn,amsmath,amssymb,aps,pre,nofootinbib]{revtex4-1}
\pdfoutput=1
\usepackage[utf8]{inputenc}
\usepackage[T1]{fontenc}
\usepackage[english]{babel}
\usepackage{graphicx}
\usepackage{bm}
\usepackage[usenames,dvipsnames]{color}
\usepackage{hyperref}
\definecolor{cobalt}{rgb}{0.0, 0.28, 0.67}
\definecolor{darkolivegreen}{rgb}{0.33, 0.42, 0.18}
\hypersetup{citecolor=darkolivegreen, linkcolor=darkolivegreen, urlcolor=cobalt, colorlinks=true}
\usepackage{pdfpages} 
\usepackage{pgffor} 

\makeatletter
\AtBeginDocument{\let\LS@rot\@undefined}
\makeatother

\def\supplementfilename{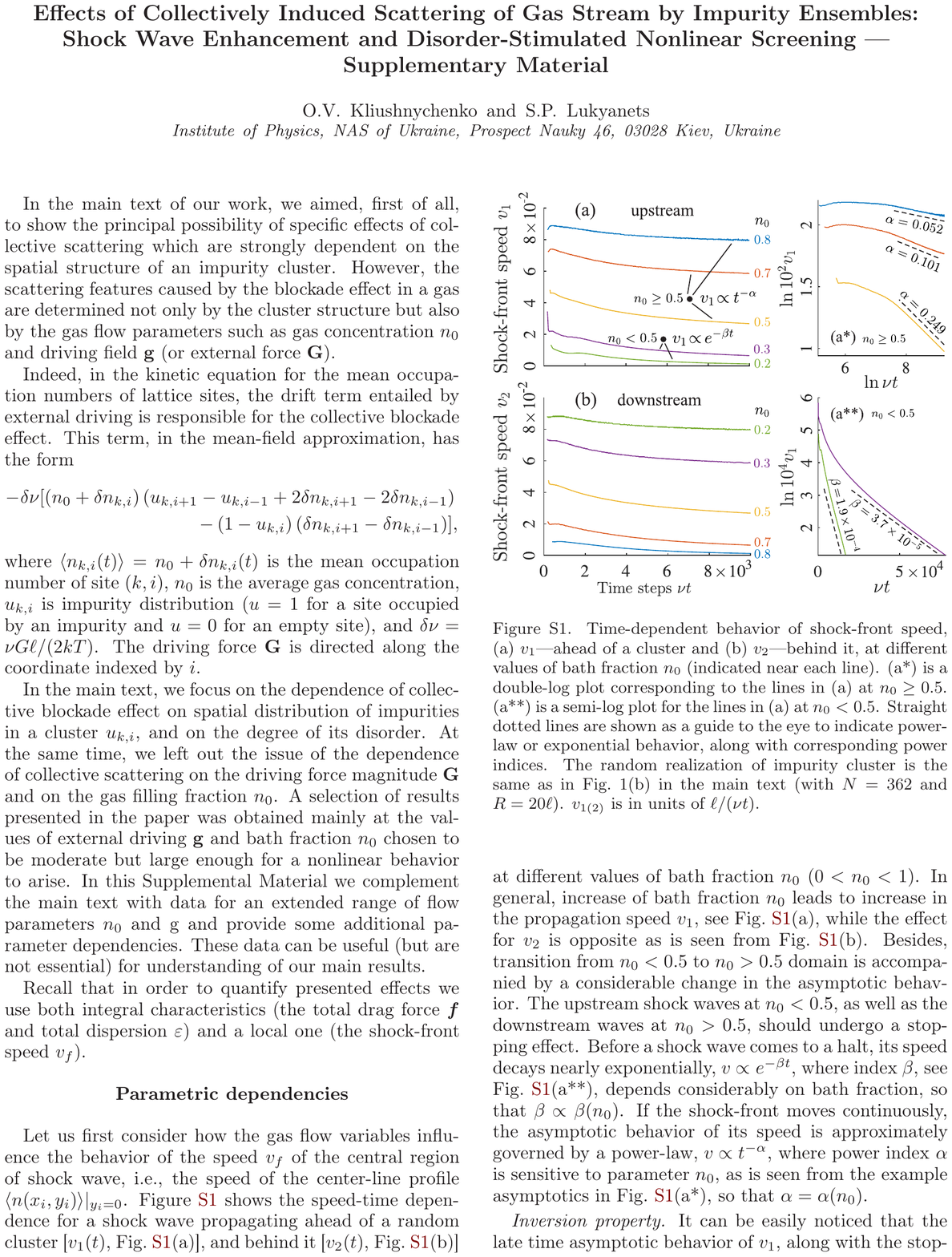}

\pdfximage{\supplementfilename}
\def\numbersupplementpages{\the\pdflastximagepages}

\newif\ifarXiv
\arXivtrue

\newcommand{\bn}{\mathbf{n}}
\newcommand{\id}{\mathrm{d}}
\newcommand{\br}{\mathbf{r}}
\newcommand{\bG}{\mathbf{G}}
\newcommand{\bg}{\mathbf{g}}
\newcommand{\bff}{\bm{f}}
\newcommand{\la}{\langle}
\newcommand{\ra}{\rangle}

\begin{document}
\title{Effects of collectively induced scattering of gas stream by impurity ensembles: Shock-wave enhancement and disorder-stimulated nonlinear screening}

\author{O.V.~Kliushnychenko} \email{kliushnychenko@iop.kiev.ua}
\affiliation{Institute of Physics, NAS of Ukraine, Prospect Nauky 46, 03028 Kiev, Ukraine}

\author{S.P.~Lukyanets} \email{lukyan@iop.kiev.ua}
\affiliation{Institute of Physics, NAS of Ukraine, Prospect Nauky 46, 03028 Kiev, Ukraine}

\begin{abstract}
We report on specific effects of collective scattering for a cloud of heavy impurities exposed to a gas stream. Formation is presented of a common density perturbation and shock waves, both generated collectively by a system of scatterers at sudden application of the stream-inducing external field. Our results demonstrate that (i) the scattering of gas stream can be essentially amplified, due to nonlinear collective effects, upon fragmentation of a solid obstacle into a cluster of impurities (heterogeneously fractured obstacle); (ii) a cluster of disordered impurities can produce considerably stronger scattering accompanied by enhanced and accelerated shock wave, as compared to a regularly ordered cluster. We also show that the final steady-state density distribution is formed as a residual perturbation left after the shock front passage. In particular, a kink-like steady distribution profile can be formed as a result of shock front stopping effect. The possibility of the onset of solitary diffusive density-waves, reminiscent of precursor solitons, is shown and briefly discussed.
\end{abstract}


\maketitle

\textit{Introduction.} Gas stream scattering by an impurity cloud often leads to pronounced collective effects. This can be manifested by formation of common perturbation ``coat'' around a cloud of scatterers or common wake, localization of gas particles (blockade effects), induced correlations and formation of nonequilibrium (dissipative) structures in the ensemble of impurities. These types of phenomena are intrinsic in various physical systems, including examples from hydrodynamics \cite{birkhoff_jets_1957,khair_motion_2007}, dusty plasmas \cite{bartnick_2016}, quantum liquids or Bose condensates \cite{pines_theory_1966,kamchatnov_stabilization_2008}, and can exhibit unusual behavior such as non-Newtonian wake-mediated forces \cite{dzubiella_depletion_2003,Sasa_2006,pinheiro_2011,Ivlev_2015,bartnick_2016,durve_2017,lisin_2017,kliushnychenko_effects_2017} that is characteristic of diffusive or dissipative systems. Spatiotemporal characteristics of medium perturbations are mostly determined by the mechanism of energy losses specific to each particular system and by the properties of medium itself, e.g., nonlinearity of associated field.

A steady-state wake profile, induced by impurities under gas flow scattering, can be considered as a residual perturbation of gas density established after its evolution over a long time. However, the properties and behavior of the system during its transit to the steady regime can significantly differ from those at steady state. For example, under abrupt activation of gas or liquid flow (or sudden impurity displacement), formation of a wake around impurities can be accompanied by propagation of a shock wave and sign change of correlation function or dissipative force between impurities \cite{frydel_long-range_2010,felderhof_long-range_2011,kliushnychenko_induced_2013}.

In this paper, we consider the properties of nonequilibrium formations resulting from scattering of gas stream by a cloud of impurities and examine the role of collective effects, with particular attention to the formation dynamics of common impurity wake (density perturbation ``coat'') in case when the stream-inducing driving field is applied suddenly (non\-adiabatically). Specifically, we analyze the properties of spatiotemporal evolution of shock waves generated collectively by a system of scatterers. We examine the effects of inner structure of impurity clusters, total drag (friction) force, and possible shock wave enhancement due to collective scattering accompanied by the nonlinear blockade effect in a gas.
Note that scattering of driven gas particles was addressed earlier (e.g., \cite{leitmann_2018}) mostly for uniform (on average) distribution of impurities throughout the entire system and in first order of the impurity density (typically $\sim10^{-3}$). Here, we consider the stream scattering on relatively dense ($\gtrsim10^{-1}$) and spatially finite clouds (clusters) of impurities.

We focus on purely dissipative (diffusive) system and make use of the minimal classical two-component lattice gas model with hard-core repulsion; that is each lattice site can be occupied by only one particle. Despite the short range of inter-particle interaction it was shown to give rise to peculiar nonlinear effects essentially manifested at high gas concentrations. These are the dissipative pairing effects \cite{kliushnychenko_effects_2017,mejia-monasterio_bias-_2011}, the wake inversion and switching of wake-mediated interaction \cite{kliushnychenko_blockade_2014,kliushnychenko_effects_2017}, formation of nonequilibrium structures \cite{benichou_2016,vasilyev_2017,poncet_2017,schmittmann_statistical_1995} etc. As will be shown, the nonlinear effects considerably affect collective scattering.

\begin{figure*}
\includegraphics[width=\textwidth]{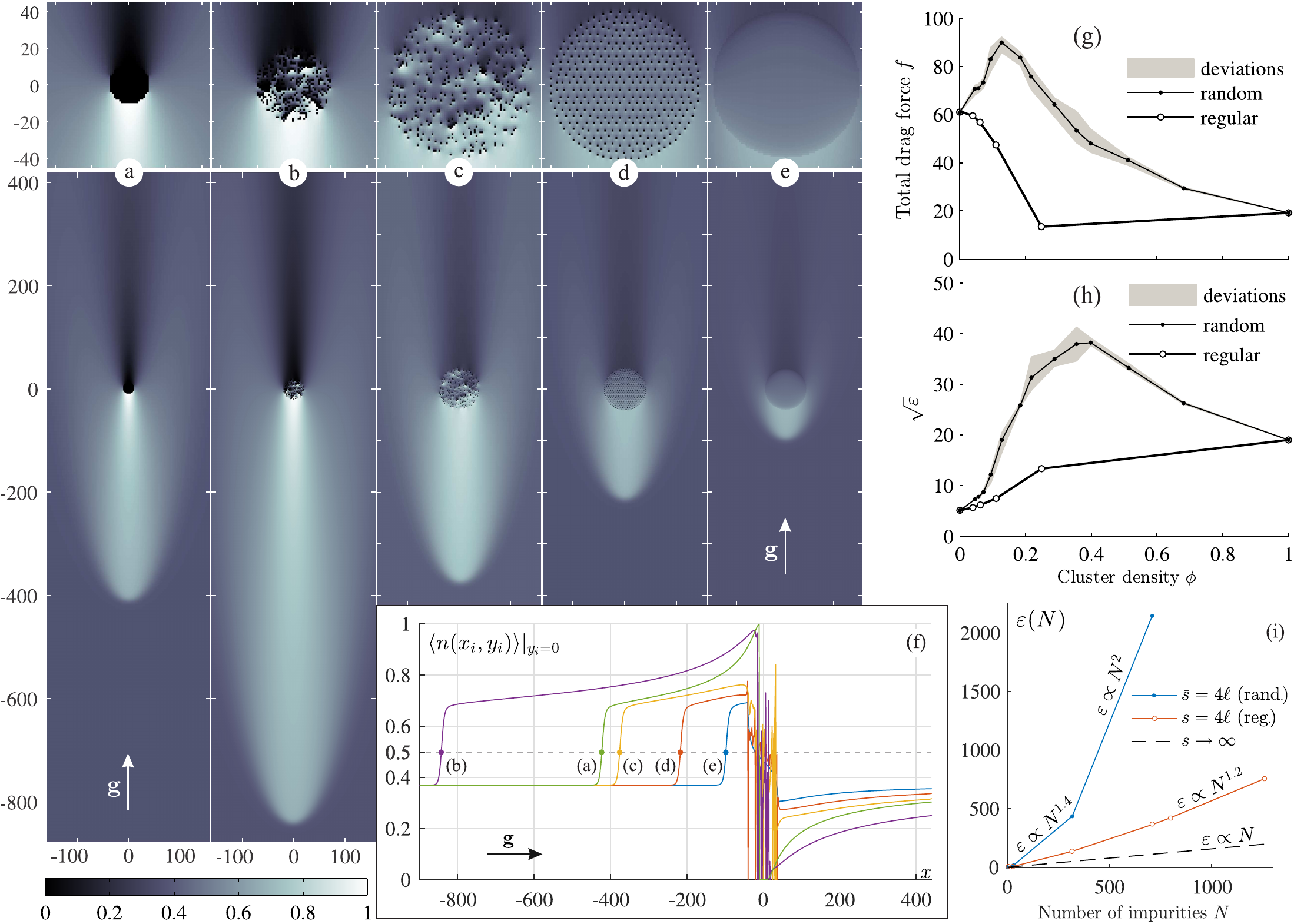}
\caption{\label{fig:cluster} (Color online) \textit{Enhancement of scattering intensity.} Steady-state distributions of mean concentration $\la n(x_i,y_i)\ra$ [panels (a) to (e)] illustrate enhanced scattering (blockade region growth) for heterogeneously fractured obstacle introduced by collective behavior. A close view of impurity cluster inner structure for each case is shown at the top: (a) solid obstacle, (b) and (c) random clusters, (d) regular cluster, (e) uniform cluster. $R=10.8\ell$ for (a), $R=20\ell$ for (b), and $R=40\ell$ for (c)--(e). Number of constituent single-site impurities is $N=362$, $n_0=0.37$, $|\bg|=0.5$ (stream is directed along the $x$-axis) for all calculated distributions. Corresponding density profiles $\la n(x_i,y_i)\ra|_{y_i=0}$ are presented on panel (f). Plots (g) and (h) show dependencies of total drag force $f\equiv|\bff|$ (units of $kT/\ell$) and $\sqrt{\varepsilon}$ on cluster density $\phi$; $R\in(\infty\,\,10.8\ell]$, $N=362$, $n_0=0.2$. Plot (i) shows dependence of dispersion $\varepsilon$ on impurity number $N$ at $n_0=0.3$; $\bar s$ (or $s$) stands for the (mean) distance between impurities, $s=4\ell$ corresponds to cluster density of $\phi=0.06$.}
\end{figure*}
Kinetics of a two-component lattice gas is described by the standard continuity equation (see, e.g., \cite{chumak_diffusion_1980,tahir-kheli_correlated_1983}), $\dot n_i^\alpha=\sum_j\left(J^\alpha_{ji}-J^\alpha_{ij}\right)+\delta J_i^\alpha$, where $\alpha=1,2$ labels the particle species and $n_i^\alpha=0,1$ are the local occupation numbers of particles at the $i$th site. $J^\alpha_{ij}=\nu^\alpha_{ij}n_i^\alpha\left(1-\sum_\beta n_j^\beta\right)$ gives the average number of jumps from site $i$ to a neighboring site $j$ per time interval, $\nu_{ij}^\alpha$ is the mean frequency of these jumps. In what follows, fluctuations of the number of jumps \cite{chumak_diffusion_1980} (the term $\delta J_i^\alpha$) are neglected. To describe the scattering of particle stream by an impurity cloud we assume, see Refs.~\cite{kliushnychenko_blockade_2014,kliushnychenko_effects_2017}, that one of the two components $u_i=0,1$ describes the given distribution of impurities and is static ($\nu_{ij}^1\equiv0$), while another one $n_i(t)$ is mobile. The presence of a weak driving field (force) $\bG$, $|\bg|=\ell|\bG|/(2kT)<1$ ($\ell$ is the lattice constant), leads to asymmetry of particle jumps for mobile component: $\nu_{ji}\approx\nu[1+\bg\cdot(\br_i-\br_j)/\ell]$. As in \cite{leung_novel_1994,schmittmann_statistical_1995,Lukyanets2010}, we use the mean-field approximation, $\partial_t\langle n_i\rangle=\sum_{j}(\langle J_{ji}\rangle-\langle J_{ij}\rangle)$, $\langle J_{ji}\rangle=\nu_{ji}\langle n_j\rangle(1-\langle n_j\rangle-u_i)$, where $\langle n_i\rangle=\langle n(\br_i)\rangle\in[0,1]$ describes the mean occupation numbers at sites $\br_i$ or the density distribution of flowing gas particles, $n_0\equiv n(|\br|\rightarrow\infty)$ being the equilibrium gas concentration (bath fraction). In what follows, we consider the two-dimensional (2D) case.

\textit{Collective scattering effects.} We start by outlining the two basic effects readily seen from Figs~\ref{fig:cluster}(a)--(e). Panels (a) to (c) represent the nonequilibrium steady-state density distribution $\la n(\br_i)\ra$ produced under scattering of streaming gas particles on the collection of point impurities arranged into the compact impermeable obstacle, dense impurity cluster (fractured obstacle) and sparse cluster, all of which consist of the same number of impurities $N$. The qualitative differences in scattered field $\delta n(\br_i)=\la n(\br_i)\ra-n_0$ constitute the essence of the first effect: Fragmentation of a solid obstacle into a cluster of separate impurities can enhance the gas stream scattering. This effect results from the collective blockade effect \cite{kliushnychenko_blockade_2014,kliushnychenko_effects_2017} which leads to screening of gas stream between impurities. For impurity cluster density $\phi=N/\pi R^2<1$ ($R$ is the cluster radius), as Fig.~\ref{fig:cluster}(b) suggests, the blockade region is considerably larger than for compact (solid) cluster [$\phi=1$, Fig.~\ref{fig:cluster}(a),(f)] as well as diluted one, $\phi\ll1$.

\begin{figure}
\includegraphics[width=\columnwidth]{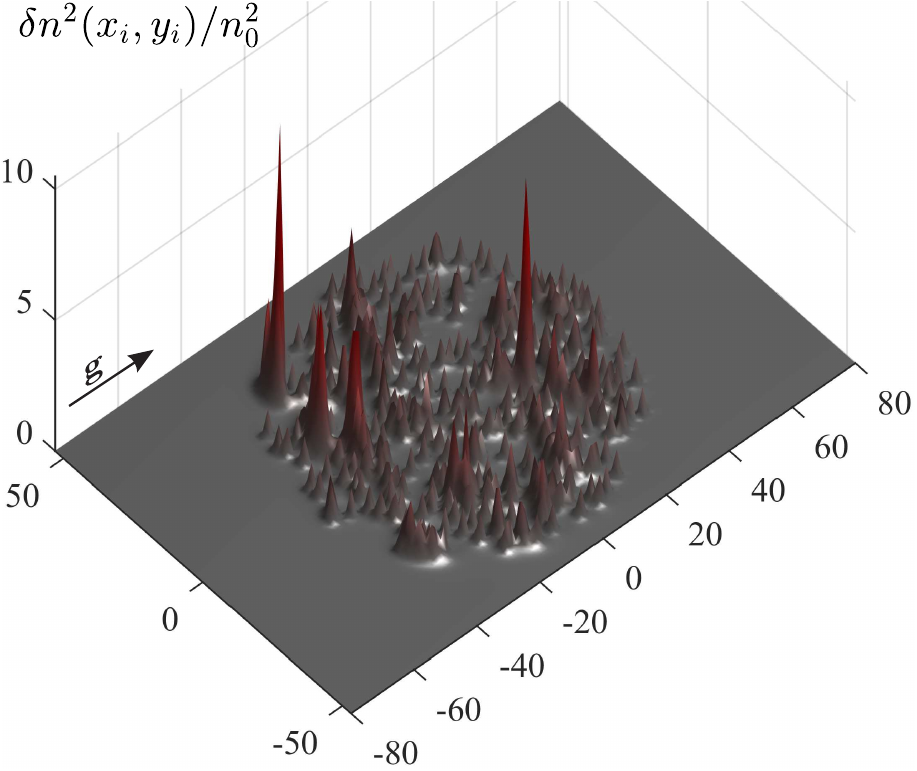}
\caption{\label{fig:peaks} (Color online) Strong local fluctuations of scattered field $\delta n(x_i,y_i)$ inside a random impurity cluster. $\phi=0.0569$ ($N=362$, $R=45\ell$), $n_0=0.2$, $|\bg|=0.5$.}
\end{figure}

The second effect consists in enhancement of scattering that is provoked by inhomogeneity of impurity distribution within a cluster, Figs~\ref{fig:cluster}(c)--(e) and (f). This effect is analogous to that of light scattering on inhomogeneities in distribution of atoms (dipole moments) that is determined by the fluctuation of their number density in a definite volume or by the two-point correlation function \cite{smoluchowski_1908,klimontovich_1949,sobelman_1971,klimontovich_1980,sobelman_2002}. As is seen from Figs~\ref{fig:cluster}(c)--(e), the scattering is less efficient for regularly ordered cluster. Note that disordered cluster can provoke strong local fluctuations of scattered field $\delta n(\br)$ inside a cluster\footnote{A similar effect of the strong local fluctuations of scattered field appears for electromagnetic field scattering on fractal clusters of nanoparticles, so-called ``hot spots'' \cite{stockman_2001,stockman_2010,sarychev1999}.},
see Fig.~\ref{fig:peaks},
i.e., $\delta n^2(\br_i)>n_0^2$. This means that the problem of gas stream scattering on such a structure cannot be adequately described by introducing the effective diffusion coefficient for a cluster or its penetration index \cite{kliushnychenko_blockade_2014}, Figs~\ref{fig:cluster}(c)--(f). In addition, this can lead to high-magnitude local fluctuations of induced dissipative interaction between impurities.\footnote{Some properties of this interaction were previously considered for a pair of impurities in \cite{kliushnychenko_effects_2017,dzubiella_depletion_2003,khair_motion_2007}.}

The magnitude of scattered field $\delta n(\br)$ can be characterized by a quantity like total density dispersion $\varepsilon\equiv\overline{\delta n^2}\propto\int\delta n^2(\br)\,\id \br$. For the cluster of $N$ infinitely distant (independent) impurities, i.e., when their mean separation length $\bar s\rightarrow\infty$, the dispersion is simply $\varepsilon\approx\sum_{i=1}^N\overline{\delta n^2_i}\approx N\overline{\delta n^2_*}\propto N$, where $\overline{\delta n^2_*}$ is dispersion for a single impurity. Figure~\ref{fig:cluster}(i) shows that dependence $\varepsilon(N)$ for impurity cluster can become power-law and, in particular, for random cluster is $\propto N^2$ that signifies the intrinsically collective scattering.

The total drag force\footnote{We exploit the drag force definition $\bff=-\int_S\bn(\br)\delta n(\br)\,\id\br$, where $\bn(\br)$ is the exterior normal of inclusion surface $S$ at the point $\br$, see \cite{dzubiella_depletion_2003,Sasa_2006,kliushnychenko_effects_2017}. For single-site inclusions we use discretized version of this expression \cite{mejia-monasterio_bias-_2011}.}
acting on impurity cluster also turns out to be quite sensitive to its density $\phi$ and cluster inner structure. As Fig.~\ref{fig:cluster}(g) suggests, the behavior is qualitatively different for random and regular clusters. At early stages of confluence, $\phi\ll1$, when collective wake-mediated interactions come into play, the regular cluster tends to reduce the total drag force exerted by the gas particles, Fig.~\ref{fig:cluster}(g). Conversely, the random cluster tends to increase the drag force
\begin{figure*}
    \includegraphics[width=\textwidth]{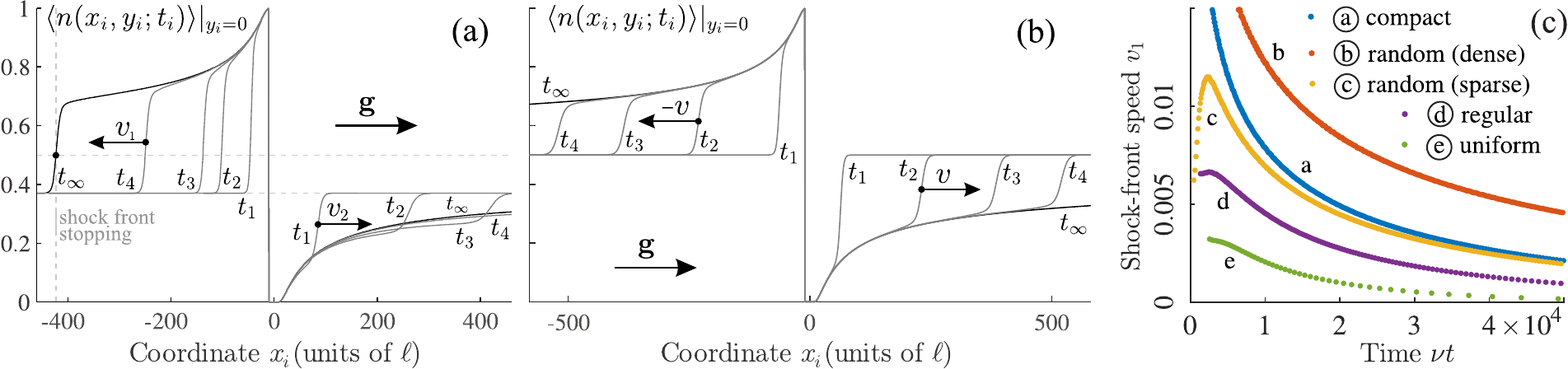}
    \caption{\label{fig:obstacle} (Color online) Density profiles $\la n(x_i,y_i)\ra|_{y_i=0}$ around impermeable obstacle of radius $R=10.8\ell$ at the time moments $t_1<t_2<t_3<t_4$ and bath fractions (a) $n_0=0.37$, and (b) $n_0=0.5$; $|\bg|=0.5$. The lines at $\nu t_\infty=5\times10^4$ show the steady-state profiles established after long-time evolution; corresponding 2D density distribution for $n_0=0.37$ (a) is shown in Fig.~\ref{fig:cluster}(a). (c) The time dependence of speed $v_1$ [in units of $\ell/(\nu t)$] of frontal shock waves propagation for clusters from Figs~\ref{fig:cluster}(a)--(e), lines are labeled to match Fig~\ref{fig:cluster}.}
\end{figure*}
until the common blockade region ahead of impurities is formed. The latter screens the impurity cluster as a whole from streaming gas particles, thereby reducing the drag force. This transformation of common perturbation ``coat’’ is reflected in the enhancement peak of the total drag force exerted on extended inhomogeneous cluster, see Fig.~\ref{fig:cluster}(g). Upon further increase of cluster density, the blockade region takes more advantageous streamlined shape, such as that shown, e.g., in  Figs~\ref{fig:cluster}(c) or (b). That also contributes towards drag force decrease until the value at $\phi=1$ (solid obstacle) is reached, Fig.~\ref{fig:cluster}(g). At the same time, the dependence of $\sqrt\varepsilon$ on random-cluster density $\phi$ has a characteristic enhancement peak which is absent for regular cluster, as shown in Fig.~\ref{fig:cluster}(h). Thus, collective gas scattering compounded by nonlinear effects can exhibit a qualitatively different behavior that depends strongly on the spatial arrangement of impurities in the cluster.

\textit{Unsteady shock-wave dynamics}. We now consider the time-dependent behavior preceding the formation of steady density distribution. Upon sudden application of stream-inducing driving field, two oppositely directed (downstream and upstream) shock waves with kink-like profiles are formed and evolve away from an impurity cluster, see Figs~\ref{fig:obstacle}(a) and (b) for the case of compact cluster (obstacle). This behavior is common for clusters of different types shown in Figs~\ref{fig:cluster}(a)--(e), in case they are large ($R\gg\ell$) and dense enough to collectively provoke nonlinear density-coat formation (for the latter to come into effect, we also should remain within a certain range of values of bath fraction and driving field \cite{sm}). The compression-like shock wave moving upstream reflects the growth dynamics of the dense region adjacent to the obstacle's surface. The rarefaction-like one in the downstream region is responsible for the formation of depleted tail or cavity (localization of vacancies). The overall stationary density profile represents the residual perturbation left after the stream scattering by impurities at $t\rightarrow\infty$ (steady scattering state). Dynamics of the two shock waves is, generally, different but obeys common inversion property upon switching from $n_0<0.5$ to $n_0>0.5$ domain \cite{sm}.

``Stopping effect''. Let us first consider formation of a dense compact region in front of a cluster, resulting from stopping of compression shock wave at $n_0<0.5$. For simplicity, we examine the behavior of center-line profile which corresponds to the central region of shock wave at $y_i=0$, Figs~\ref{fig:obstacle}(a) and (b), for the case of impermeable circular obstacle. As can be seen from Fig.~\ref{fig:obstacle}(a), the shock wave comes to a halt at certain standoff distance $x_f^*$ ($x_f$ is the front\footnote{Shock front position $x_f$ corresponds to the inflexion point of $\la n(x_i,y_i)\ra|_{y_i=0}$. As $x_f$ takes discrete values, we use the smoothing procedure \cite{sm} for $x_f(t)$ and $v_f(t)$.} position):
once the condition $n_f=n(x_f,0)=0.5$ is reached, the motion vanishes, i.e., shock-front speed $v_f\rightarrow0$. This criterion holds also for non-regular clusters, see Fig.~\ref{fig:cluster}(f). The front speed asymptotically decays as $v_f(t)\propto e^{-\gamma t}$. The front speed of downstream shock wave decays according to power-law and does not undergo stopping, forming a depleted tail (for details see \cite{sm}).

At $n_0>0.5$, there is no stopping effect for the upstream shock wave, while it does occur for the downstream wave. Such ``switching'' is in agreement with the wake-inversion effect \cite{kliushnychenko_blockade_2014,kliushnychenko_effects_2017,sm}. At $n_0=0.5$, both shock waves do not undergo stopping and decay asymptotically as $n_f\propto 1/2\pm At^{-\beta}$, Fig.~\ref{fig:obstacle}(b).

Shock-front stopping criterion can be roughly estimated within the continual approximation \cite{kliushnychenko_blockade_2014,kliushnychenko_effects_2017}. Coarse description of the system is given by the Burgers-type equation that admits a kink-like solution. Far from the cluster, this equation takes the form $\partial_t n=\nabla^2n-(\bg\cdot\nabla)[n(1-n)]$. The growth of the blockade region (shock front propagation) takes place when inflow of the gas particles prevails over their outflow via lateral diffusion. One may suppose that, for large and dense impurity cluster, accumulation dynamics in the central region (center-line at $y=0$) can be approximately described by a quasi-1D equation. Its solution represents a kink approaching values $n_-$, $n_+$ as $x\rightarrow\pm\infty$ and has the form $n(x,t) = n_0-\Delta n\tanh{g\Delta n[x+v_ft-x_0]}$, $\Delta n=|n_--n_+|/2$ is the shock half-height and the speed of shock wave $v_f$ is given by $v_f=2g\left(1/2-n_f\right)$, where $n_f=(n_-+n_+)/2$.\footnote{$n_\pm\approx n(x_f\pm\Delta)$ can be associated with shock height, $\Delta$ is the shock front thickness.} At $n_f\rightarrow0.5$, the shock front motion vanishes, i.e., $v_f\rightarrow0$. This stands for quantitative stopping criterion which is in agreement with numerical results for the center-line profile of shock wave in the 2D case, Figs~\ref{fig:obstacle}(a), \ref{fig:cluster}(f). Quasi-1D case allows only to estimate the stopping condition and does not give full description of 2D shock wave, its lateral region, asymptotic behaviors.

Now we consider the effects caused essentially by irregularity in the distribution of impurities. The first specific property is that decomposition of solid obstacle or homogeneous impurity cluster into inhomogeneous cluster can result in the enhancement of shock wave---increase in its amplitude and front speed, see Fig.~\ref{fig:obstacle}(c) and corresponding steady-state profiles in Fig.~\ref{fig:cluster}. The other property is what can be referred to as precursor effect.

\begin{figure}
\includegraphics[width=\columnwidth]{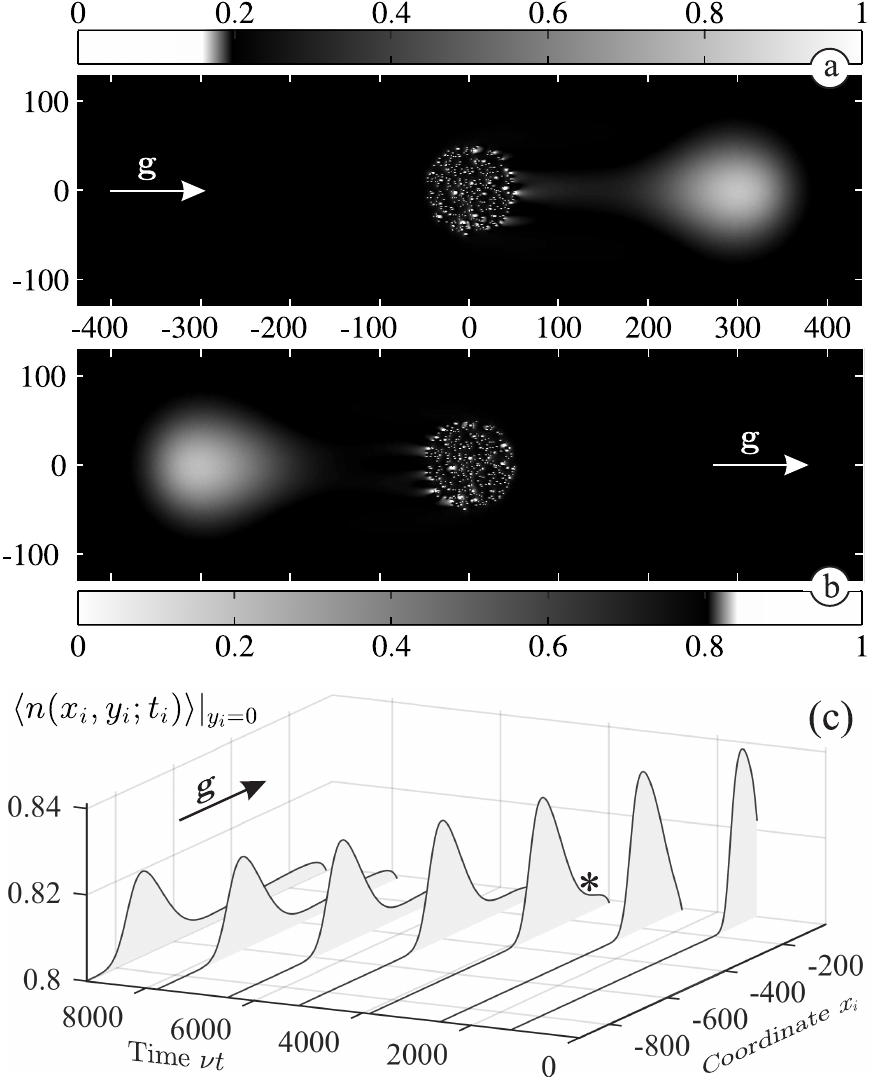}
\caption{\label{fig:hump} Momentary gas distributions $|\delta n(x_i,y_i)|$ at $\nu t=3.9\times10^3$ illustrating the precursor propagation: bunch of gas particles (b) or holes (a) separating from the impurity cluster. (c) The time evolution of density profile $\delta n(x_i,\nu t;y_i=0)$; the profile marked by an asterisk corresponds to panel (b). Cluster density $\phi=0.0461$, $n_0=0.2$ for (a), $n_0=0.8$ for (b) and (c), $|\bg|=0.5$.}
\end{figure}

At the initial stage, just after abrupt application of the external drive, an impurity cluster is capable to generate larger density perturbation, both ``stored'' within a cluster and around it, than that hold in a steady state. For this reason, the system tends to subsequently get rid of the excess density perturbation that can be realized by two mechanisms:

(i) The excessive portion of density perturbation ``leaves'' the cluster in form of solitary wave traveling downstream\footnote{This recalls the avalanche problem for a pile of sand \cite{bak_1996}.} at $n_0<0.5$, Fig.~\ref{fig:hump}(a),
or upstream at $n_0>0.5$, Figs~\ref{fig:hump}(b) or (c), that is in accordance with wake inversion effect \cite{kliushnychenko_blockade_2014}. The ``ejected'' bunch of gas particles (or vacancies) and associated hump-shaped wave (or the inverse hump) have a characteristic length-scale (its half-width) commensurate with the cluster size $2R$. This precursor-like mechanism takes place for sparse clusters, $\phi\ll1$, for which the common blockade region ahead of scatterers is either weak or not formed.

\begin{figure}
\includegraphics[width=\columnwidth]{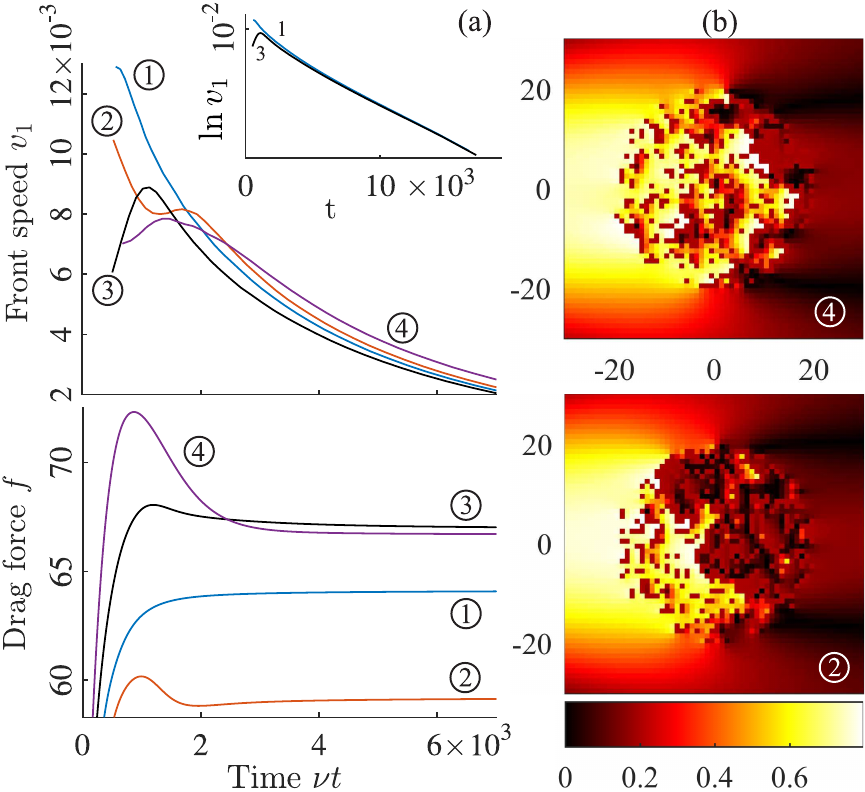}
\caption{\label{fig:reals} (Color online) Time dependence of the shock-front speed and total drag force (a) for several realizations 1--4 of random impurity-cluster. $N=362$, $R=20\ell$, $|\bg|=0.5$, $n_0=0.2$, $f$ and $v_1$ are in units of $kT/l$ and $\ell/(\nu t)$, correspondingly. The front speed slowdown is nearly exponential, $\ln v_1\propto-t$, see the inset in (a). (b) Steady-state distributions of $\delta n^2(x_i,y_i)$ within a cluster, realizations 4, 2.}
\end{figure}

(ii) A different mechanism comes into play for modestly dense clusters, when the excessive perturbation is relaxed via temporal acceleration of the shock front at the initial times. It can be explained by considering the dynamic behavior for several realizations of random cluster with an intermediate value of fraction $\phi\approx10^{-1}$, see Fig.~\ref{fig:reals}.
As Fig.~\ref{fig:reals}(a) suggests, shock front speed acceleration is sensitive to the realizations of random cluster. This temporal acceleration is always preceded with or accompanied by the enhancement peak of total drag force, Fig.~\ref{fig:reals}(b), while for the cluster realization without acceleration effect the drag force exhibit monotonic saturation. The temporal force enhancement signifies the presence of excess perturbation (within a cluster) that subsequently transfers into shock wave acceleration. Under sudden stream activation, the nonlinear blockade effect leads to local saturation of scalar density field $n(\br,t)$ attained faster than the overall perturbation is redistributed to minimize the total drag force. This leads to the accelerated growth of blockade region at the initial times.

Note, system behavior is also sensitive to the variables of gas flow: bath fraction $n_0$ and magnitude of driving force $\bG$. Shock front stopping property is consistent with concentration-dependent wake-inversion effect of \cite{kliushnychenko_blockade_2014}. In the domain $n_0<0.5$, increasing of $n_0$ and/or $\bG$ leads to the enhancement of scattering and shock wave, while qualitative picture of scattering is not changed, so we give details on this study in Supplemental Material \cite{sm}.

\textit{Concluding remarks}.
We presented some results on the effects of collective scattering of gas stream on finite-sized impurity clouds and associated shock-wave generation, both accompanied by a nonlinear blockade effect. Based on a numerical solution of kinetic equations for the average local occupation numbers of lattice gas, we display significant role of spatial disorder of scatterers (impurities). Our results show that scattering of gas flow on impurity cloud and shock-wave generation can be enhanced by decomposition of solid obstacle into fragments or a sparse cluster of impurities. This enhancement is more efficient for disordered clusters as compared to regular ones. Note that the shock wave amplification effect correlates, to a certain extent, with the well known problem of air shock wave interaction with a porous screen where the effect of temporal enhancement of reflected shock wave was observed \cite{gelfand_1975,gelfand_1983}. In addition, disordered cluster of scatterers can provoke high local fluctuations of scattered field inside the cluster and avalanche-like effect at sudden application of the external driving field. Considered effects reveal a close formal analogy to classical problem of light scattering in atomic, molecular, or nanoparticle ensembles.

The simplest hard-core lattice gas model alone leads to peculiar nonlinear effects mentioned above, which can be of interest considering kinetics of adatoms on solid surfaces \cite{chumak_diffusion_1980}, surface electromigration \cite{schimschak_1997}, or superionic conductors \cite{schmittmann_statistical_1995}. Further application of more realistic Coulomb, Yukawa, or Lennard-Jones potentials of interparticle interaction, can better describe particular physical systems, e.g., dusty plasmas or colloidal dispersions, see \cite{tsytovich_nonlinear_2013,tsytovich_self_2015,sriram_out--equilibrium_2012,vasilyev_2017}.

\textit{Acknowledgments.} We are grateful to V.V. Gozhenko, Prof. B.I. Lev, and E.V. Stolyarov for their attention to this work.

\bibliography{refs}

\ifarXiv
    

\section*{Supplementary material}

# Effects of Collectively Induced Scattering of Gas Stream by Impurity Ensembles: Shock Wave Enhancement and Disorder-Stimulated Nonlinear Screening — Supplementary Material

O.V. Kliushnychenko and S.P. Lukyanets

*Institute of Physics, NAS of Ukraine, Prospect Nauky 46, 03028 Kiev, Ukraine*

In the main text of our work, we aimed, first of all, to show the principal possibility of specific effects of collective scattering which are strongly dependent on the spatial structure of an impurity cluster. However, the scattering features caused by the blockade effect in a gas are determined not only by the cluster structure but also by the gas flow parameters such as gas concentration $n_0$ and driving field $\mathbf{g}$ (or external force $\mathbf{G}$).

Indeed, in the kinetic equation for the mean occupation numbers of lattice sites, the drift term entailed by external driving force is responsible for the collective blockade effect. This term, in the mean-field approximation, has the form

$$-\delta\nu[(n_0 + \delta n_{k,i})(u_{k,i+1} - u_{k,i-1} + 2\delta n_{k,i+1} - 2\delta n_{k,i-1})$$
$$- (1 - u_{k,i})(\delta n_{k,i+1} - \delta n_{k,i-1})],$$

where $\langle n_{k,i}(t)\rangle = n_0 + \delta n_{k,i}(t)$ is the mean occupation number of site $(k, i)$, $n_0$ is the average gas concentration, $u_{k,i}$ is impurity distribution ($u = 1$ for a site occupied by an impurity and $u = 0$ for an empty site), and $\delta\nu = \nu G\ell/(2kT)$. The driving force $\mathbf{G}$ is directed along the coordinate indexed by $i$.

In the main text, we focus on the dependence of collective blockade effect on spatial distribution of impurities in a cluster $u_{k,i}$, and on the degree of its disorder. At the same time, we left out the issue of the dependence of collective scattering on the driving force magnitude $\mathbf{G}$ and on the gas filling fraction $n_0$. A selection of results presented in the paper was obtained mainly at the values of external driving $\mathbf{g}$ and bath fraction $n_0$ chosen to be moderate but large enough for a nonlinear behavior to arise. In this Supplemental Material we complement the main text with data for an extended range of flow parameters $n_0$ and g and provide some additional parameter dependencies. These data can be useful (but are not essential) for understanding of our main results.

Recall that in order to quantify presented effects we use both integral characteristics (the total drag force $\boldsymbol{f}$ and total dispersion $\varepsilon$) and a local one (the shock-front speed $v_f$).

## Parametric dependencies

Let us first consider how the gas flow variables influence the behavior of the speed $v_f$ of the central region of shock wave, i.e., the speed of the center-line profile $\langle n(x_i, y_i)\rangle|_{y_i=0}$. Figure S1 shows the speed-time dependence for a shock wave propagating ahead of a random cluster [$v_1(t)$, Fig. S1(a)], and behind it [$v_2(t)$, Fig. S1(b)]

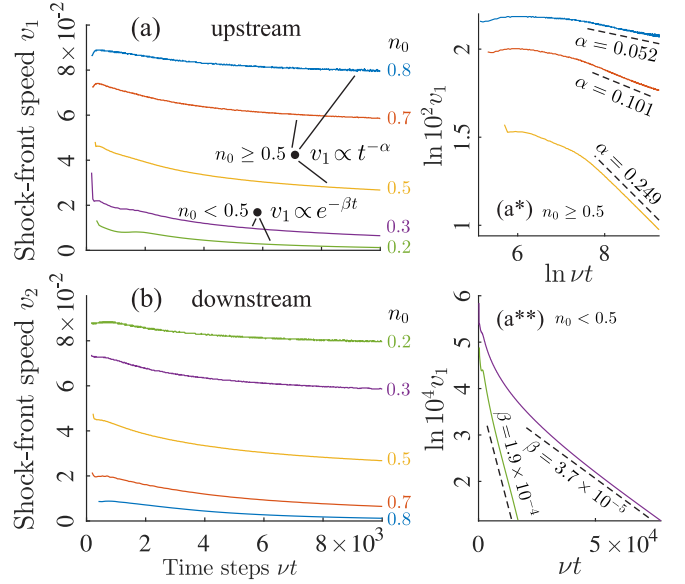

Figure S1. Time-dependent behavior of shock-front speed, (a) $v_1$—ahead of a cluster and (b) $v_2$—behind it, at different values of bath fraction $n_0$ (indicated near each line). (a*) is a double-log plot corresponding to the lines in (a) at $n_0 \geq 0.5$. (a**) is a semi-log plot for the lines in (a) at $n_0 < 0.5$. Straight dotted lines are shown as a guide to the eye to indicate power-law or exponential behavior, along with corresponding indices. The random realization of impurity cluster is the same as in Fig. 1(b) in the main text (with $N = 362$ and $R = 20\ell$). $v_{1(2)}$ is in units of $\ell/(\nu t)$.

at different values of bath fraction $n_0$ ($0 < n_0 < 1$). In general, increase of bath fraction $n_0$ leads to increase in the propagation speed $v_1$, see Fig. S1(a), while the effect for $v_2$ is opposite as is seen from Fig. S1(b). Besides, transition from $n_0 < 0.5$ to $n_0 > 0.5$ domain is accompanied by a considerable change in the asymptotic behavior. The upstream shock waves at $n_0 < 0.5$, as well as the downstream waves at $n_0 > 0.5$, should undergo a stopping effect. Before a shock wave comes to a halt, its speed decays nearly exponentially, $v \propto e^{-\beta t}$, where index $\beta$, see Fig. S1(a**), depends considerably on bath fraction, so that $\beta \propto \beta(n_0)$. If the shock-front moves continuously, the asymptotic behavior of its speed is approximately governed by a power-law, $v \propto t^{-\alpha}$, where power index $\alpha$ is sensitive to parameter $n_0$, as is seen from the example asymptotics in Fig. S1(a*), so that $\alpha = \alpha(n_0)$.

*Inversion property.* It can be easily noticed that the late time asymptotic behavior of $v_1$, along with the stop-

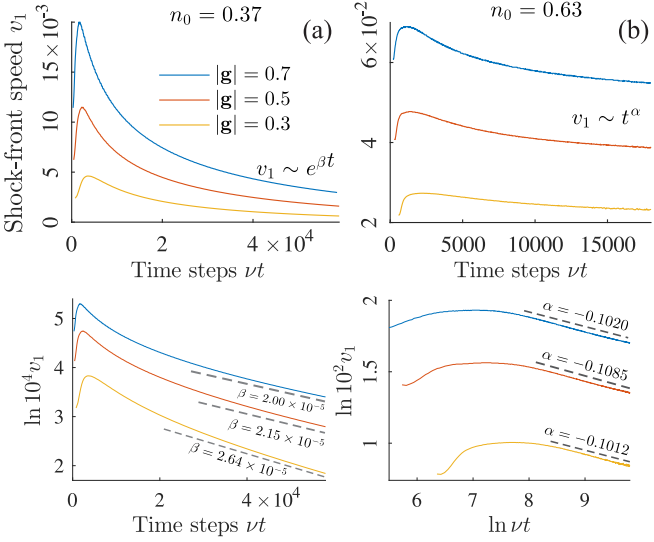

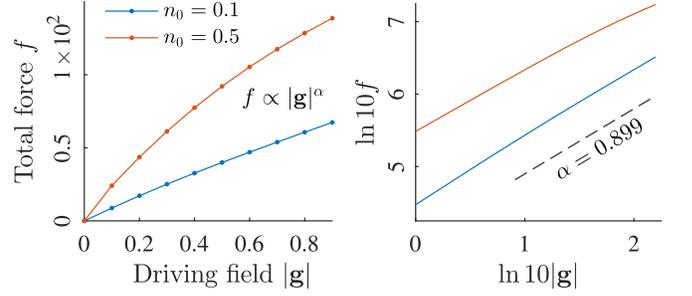

Figure S4. The total drag force acting on a random cluster in a steady-state case. Left: The total force $f \equiv |\boldsymbol{f}|$ (in units of $kT/\ell$) against the driving field $|\mathbf{g}|$, at two values of bath fraction $n_0$. Right: Corresponding log-log plot, showing typical asymptotic behavior. The random realization of impurity cluster is the same as in Fig. 1(c) in the main text (with $N = 362$ and $R = 40\ell$).

Figure S2. Enhancement of shock-front speed $v_1$ (ahead of cluster) at increasing values of driving field $|\mathbf{g}|$ for (a) $n_0 = 0.37$, and (b) $n_0 = 0.63$. Below each of the panels, corresponding log-log and semi-log dependencies are shown with approximate asymptotics. The random realization of impurity cluster is the same as in Fig. 1(c) in the main text (with $N = 362$ and $R = 40\ell$). $v_1$ is in units of $\ell/(\nu t)$.

ping effect, at each value of $n_0$ is similar to that of $v_2$ at bath fractions $1 - n_0$. This reflects the property of concentration-dependent wake inversion [1] (see, e.g., Fig. 1 from Ref. [1] or Figs 1 and 5 from Ref. [2]).

Dependence of shock-front speed $v_1$ on the strength of driving field $\mathbf{g}$ is illustrated in Fig. S2 for two values of $n_0$, one chosen from $n_0 < 0.5$ domain, Fig. S2(a), and another one from $n_0 < 0.5$, Fig. S2(b). In the first case, the late time slowing dynamics is characterized by exponential decay of shock front speed towards full stopping,

$v_1 = 0$, while for the second case, decay is approximately governed by a power-law dependence. In the latter case, power index $\alpha < 1$ remains almost unaffected with increasing driving field magnitude, see Fig. S2(b*). For the exponential decay, Fig. S2(a*), there is only a slight decrease of index $\beta$.

The force exerted on random impurity cluster at varying bath fraction $n_0$ is shown in Fig. S3. $f(t)$ typically undergoes the enhancement peak at initial times and quite quickly comes to its stationary value $f(t \to \infty)$. In turn, steady-state values of $f(n_0)$ lie on the flattened-hump curve which is almost symmetrical, again, due to the concentration-dependent wake-inversion property [1, 2], see inset in Fig. S3. This lengthy flat region reflects the presence of intense particle flow screening and efficient cluster streamlining, the total force is saturated at lower values of $n_0$ as compared to the case of a single impurity or a compact obstacle (see [2]).

The influence of driving field strength on the total drag force is presented for steady-state case, see Fig. S4. As seen, the force saturation, with increasing driving field $|\mathbf{g}|$, is generally governed by a power law $\sim A|\mathbf{g}|^\alpha$, where $\alpha \lesssim 1$. As the inset in Fig. S4 shows, the power index $\alpha$ depends weakly on $n_0$, i.e., its asymptotic behavior is approximately the same for various values of bath fraction, while the "amplitude" $A = A(n_0)$ is a nonmonotonic function of $n_0$.

Increasing of bath fraction $n_0$ also gives rise to enhancement of scattering quantified by total dispersion $\varepsilon$, see Fig. S5. Notice that, in the particular case of $n_0 = 0.5$, the time-saturation of quantity $\sqrt{\varepsilon}$ is governed by a power law $\propto t^\alpha$ ($\alpha < 1$), see Fig. S5(Right). Moreover, this asymptotic behavior holds at various values of the driving field $|\mathbf{g}|$, the value of power index $\alpha$ being modified slightly but not substantially, see Fig. S6(Right).

As a general conclusion, in the domain $n_0 < 0.5$, increasing of $n_0$ and/or $\mathbf{G}$ leads to enhancement of both the scattering and the shock wave without changes in the qualitative picture of scattering.

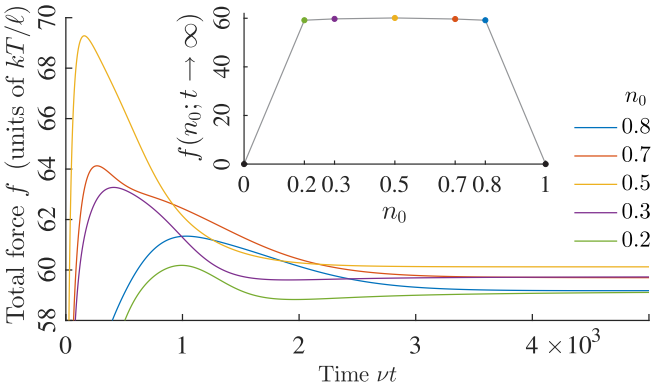

Figure S3. The total force $f = |\boldsymbol{f}|$, acting on impurity cluster, as a function of time at different values of bath fraction $n_0$. The inset: the steady-state values of $f$ in the full range of $n_0$ values. The random realization of impurity cluster is the same as in Fig. 1(b) in the main text (with $N = 362$ and $R = 20\ell$).

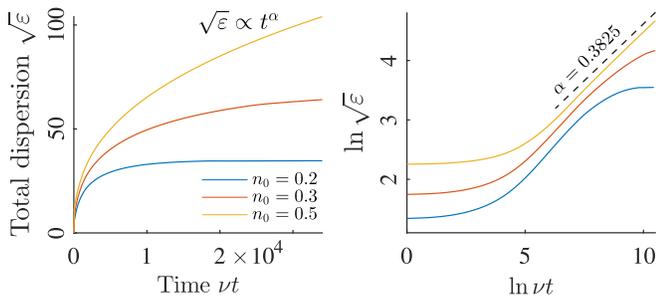

Figure S5. Left: The time dependence of total dispersion $\sqrt{\varepsilon}$ at different values of gas concentration $n_0$, $|\mathbf{g}| = 0.5$. Right: Corresponding log-log plot, showing typical asymptotic behavior. The random realization of impurity cluster is the same as in Fig. 1(c) in the main text (with $N = 362$ and $R = 40\ell$).

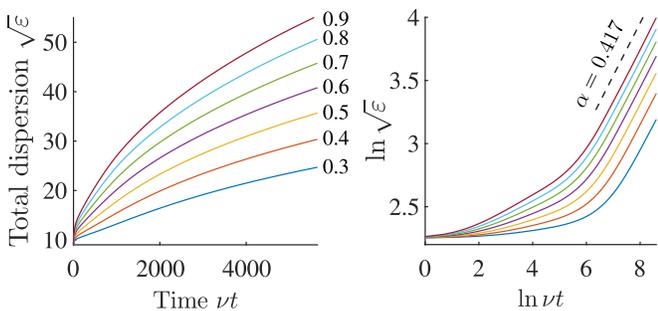

Figure S6. Left: The time dependence of total dispersion $\sqrt{\varepsilon}$ at different values of driving field $|\mathbf{g}|$ (indicated near each line), $n_0 = 0.5$. Right: Corresponding log-log plot, showing typical asymptotic behavior. The random realization of impurity cluster is the same as in Fig. 1(c) in the main text (with $N = 362$ and $R = 40\ell$).

## A note on front dynamics

It was briefly mentioned in the main text that position of the shock front is defined by the site $x_f$ for which the second derivative of $\langle n(x_i, y_i) \rangle|_{y_i=0}$ (finite-difference derivative on a lattice) is zero. For this reason, $x_f$ can take only discrete values along the center-line density-profile. Also, within this approach, the time-dependence $x_f(t)$, as well as $v_f(t)$, is not smooth and reflects inherent spatial discreteness of the system. In particular, the motion of shock front exhibits a "stair-step" behavior as shown in Fig. S7(a). Such behavior is often encountered in spatially discrete systems, e.g., sequential overcoming of barriers in diffusive me-

dia with strong inhomogeneity [3] or so-called "halt-and-go" type dynamics of fronts caused by lattice and finite particle effects [4]. To evaluate $v_f(t)$ (as a finite-difference derivative) properly, obtained nonsmooth dependence $x_f(t)$ requires a sort of regularization procedure, so that we replace it with a "smooth" piecewise linear function giving averaged behavior as shown in Fig. S7(a). Note that evaluated function $v_f(t)$ can be also non-monotonic with small-scale quasi-periodic oscillations as shown in Fig. S7(b). An approximated smoothed fit can be roughly obtained, e.g., by simply using the central difference defined on next-nearest-neighbour sites $v_f(t_i) \approx [x_f(t_{i+k\ell}) - x_f(t_{i-k\ell})]/\Delta t$; the larger the values of $k = 1, 2, \ldots$, the more smoothed curve we get.

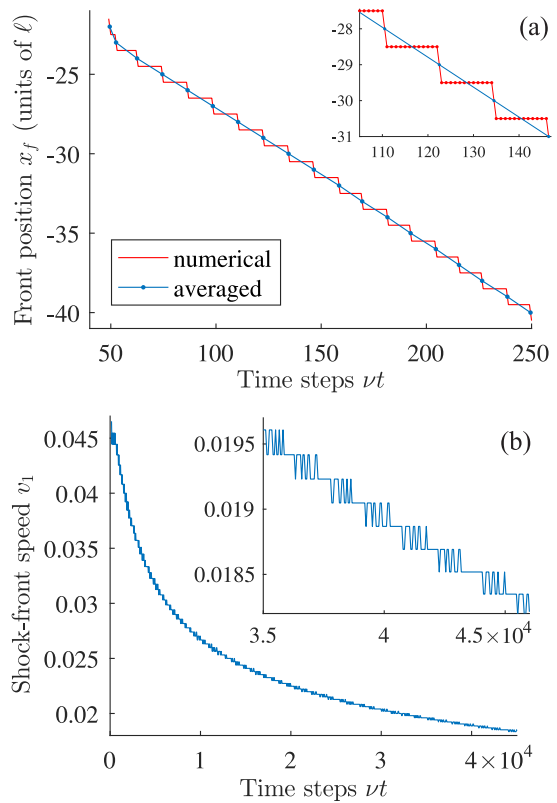

Figure S7. (a) $x_f(t)$ regularization. The time dependence of upstream shock-front $x_f$ at bath fraction $n_0 = 0.8$. (b) The time dependence of upstream shock-front speed $v_1$ (in units $\ell/\nu t$) at bath fraction $n_0 = 0.5$. The random realization of impurity cluster is the same as in Fig. 1(b) in the main text (with $N = 362$ and $R = 20\ell$). External field $|\mathbf{g}| = 0.5$.


\fi

\end{document}